\def\year{2019}\relax
\documentclass[letterpaper]{article} %DO NOT CHANGE THIS
\usepackage{aaai19}   %Required
\usepackage{times}    %Required
\usepackage{helvet}   %Required
\usepackage{courier}  %Required
\usepackage{url}
\usepackage{graphicx} %Required
\title{Demographic Differentials in Facebook Usage Around the World}
\author{
Sofia Gil-Clavel,
Emilio Zagheni\\
Max Planck Institute for Demographic Research\\
gil@demogr.mpg.de, zagheni@demogr.mpg.de
}

\begin{document}

\maketitle

\begin{abstract}
We use data from the Facebook Advertisement Platform to study patterns of demographic disparities in usage of Facebook across countries. We address three main questions: (1) How does Facebook usage differ by age and by gender around the world? (2) How does the size of friendship networks vary by age and by gender? (3) What are the demographic characteristics of specific subgroups of Facebook users? We find that in countries in North America and northern Europe, patterns of Facebook usage differ little between older people and younger adults. In Asian countries, which have high levels of gender inequality, differences in Facebook adoption by gender disappear at older ages, possibly as a result of selectivity. We also observe that across countries, women tend to have larger networks of close friends than men, and that female users who are living away from their hometown are  more likely to engage in Facebook use than their male counterparts, regardless of their region and age group. Our findings contextualize recent research on gender gaps in online usage, and offer new insights into some of the nuances of demographic differentials in the adoption and the use of digital technologies. 

\end{abstract}

\section{Introduction}
In recent years, there has been increasing interest in understanding how people's use of Information and Communication Technologies (ICTs) is related to their individual characteristics such as their age, gender, education, mental health, and income. Studies have focused not only on the determinants of ICT use, but on the implications that ICTs have for people's everyday lives. Whether the effects of ICTs are mostly positive or mostly negative is still unclear, but it is often observed that ICTs can provide people with emotional support, help them maintain old friendships or form new ones, and provide access to relevant information~\cite{neves_does_2015}. While efforts to study the reasons behind ICT adoption have been made, we still lack a full understanding of what motivates different population groups across the world to use ICTs \cite{blaschke_ageing_2009}. 

Earlier research on ICT use mostly focused on young populations, whereas the few studies that focused on ICT among the elderly were mainly carried out in developed countries~\cite{jimison_barriers_2008,helsper_ageing_2009,czaja_factors_2006}. Given that the elderly are more likely to suffer from social isolation and health problems, and that population aging is becoming a global trend and a major challenge for our societies, studying older population age groups is increasingly important. Recent literature has provided evidence that problems like social isolation can be alleviated through the use of ICTs~\cite{jimison_barriers_2008,bradley_assistive_2003}. However, despite the worldwide spread of digital technologies and the increase in the use of social networking sites, it appears that older population age groups remain largely disengaged from ICTs \cite{neves_does_2015}. 

In this paper, we leverage a relatively untapped source, Facebook (FB) data for advertisers, to examine demographic disparities in FB adoption. We consider these data for at least three main reasons: (i) as FB is the most widely used social  networking site on a global scale, the prevalence of FB use can be considered as a first approximation for access to digital technologies~\cite{fatehkia_using_2018,garcia_analyzing_2018}; (ii) the FB advertisement platform allows users to query counts of users broken down by demographic characteristics like age, gender, and interests; and (iii) the prevalence of FB use has been studied at a global scale only with respect to gender gaps~\cite{fatehkia_using_2018,garcia_analyzing_2018} and migration~\cite{zagheni_leveraging_2017,dubois_studying_2018}, whereas research on `age gaps', as well as on differential usage of the platform broken down by broad demographic characteristics, is missing.    

In short, in this paper we address three main questions: (1) How does FB use differ by age and by gender around the world? (2) How does the size of friendship networks vary by age and gender? (3) What are the demographic characteristics of specific subgroups of FB users? 

Addressing these questions would essentially tell us how the prevalence of FB usage varies according to demographic characteristics. Analyzing this information is important for understanding gaps in the adoption of digital technologies across demographic groups and is a first step towards assessing the effects of disparities in social media use on well-being.

\section{Data}

FB has developed an Advertisement Platform that allows advertisers to target specific  populations they would like to reach. Before ads are created and the advertisers incur costs, FB produces estimates of the target audience, which are typically used by the advertisers for budgetary purposes. These data can be retrieved programmatically via an API. For detailed information about the data and how they can be accessed, see Zagheni et al. (2017). 

For this paper, we retrieved two different types of data: first, the total estimated daily number of FB or Messenger users; second, the estimated daily number of FB or Messenger users that match certain demographic and behavioral characteristics. Both databases are disaggregated by age and gender for 232 countries. Since FB estimates of the daily number of FB users can be noisy, we collected data for the same variables on six different days during the week of November 13-19, 2018. We also considered relatively broad age groups defined as follows: Young Adulthood (16-24 years), Middle Adulthood (25-54 years), and Late Adulthood (55-65 years).  We then used the median value for each variable across all six data collections as a summary measure of the central tendency in each country considered.

Small countries or countries not supported by FB for which the value of users in the Late Adulthood group was consistently equal to zero (e.g. the median was equal to zero) were omitted from our analysis. The final database included data on 136 countries, disaggregated by gender, age group, and selected behavioral interests. For this work, the African countries were not considered, given that the data showed a huge amount of noise. China was also not considered due to the restrictions in FB use in the country. Finally,  Cuba, Iran, North Korea, Sudan, and Syria did not appear in our data base, as in those countries FB cannot post ads by law.

\section{Methods}

We propose a series of indexes to summarize the information available in our database and to present relative differences in the prevalence of FB use across demographic characteristics. We define the ratio $R(i,j)_{s,c}$ of FB users of gender $s$ between ages $i$ and $j$ in country $c$ as:
\begin{equation}\label{EqRatios}
R(i,j)_{s,c} =\frac{FB(i,j)_{s,c}}{Pop(i,j)_{s,c}}    
\end{equation}
where $FB(i,j)_{s,c}$ is the population size of gender $s$ between ages $i$ and $j$ from country $c$, and $Pop(i,j)_{s,c}$ is the total population size of sex $s$ between ages $i$ and $j$ from country $c$ (Figure 1). In order to obtain the total number of people between the ages $i$ and $j$ for 2018, we computed a linear approximation using the 2015 and 2020 data from the United Nations World Population Prospects. 

As one of our goals is to compare how relative adoption of FB changes between the Middle and the Late Adulthood groups, we propose a summary index that we call the `Facebook Elderly Penetration Index' (FEPI), and that we define as follows: 
\begin{equation}\label{EqFEPI}
FEPI_{s,c}=\frac{R( 55,65 )_{s,c}}{R( 25,54 )_{s,c}}  
\end{equation}
This index is useful for comparing the number of older people who are using FB with the number of people in middle adulthood in relative terms, while accounting for the fact that overall FB penetration rate varies across countries. A FEPI value below one indicates that in relative terms, the older adults who reside in country $c$ use FB less than the middle-aged adults.

A second goal of this paper is to evaluate  the prevalence of subgroups of FB users with certain specific characteristics, such as friendship network size. We define the following index $I$ as: 
\begin{equation}\label{EqIndex}
I(i,j)_{s,c}[variable]=\frac{FB(i,j)[variable]_{s,c}}{FB(i,j)_{s,c}}
\end{equation}
Where $FB(i,j)_{s,c}[variable]$ is the total number of FB users of gender $s$ between ages $i$ and $j$, who live in country $c$ and match the characteristic $variable$ (provided by FB).
$FB(i,j)_{s,c}$ is the total number of FB users of gender $s$ between ages $i$ and $j$, who live in country $c$.

\section{Results}

\subsection{Disparities in FB Adoption by Age and by Gender}

\begin{figure}[ht]
    \begin{center}
    \includegraphics[width=.95\columnwidth]{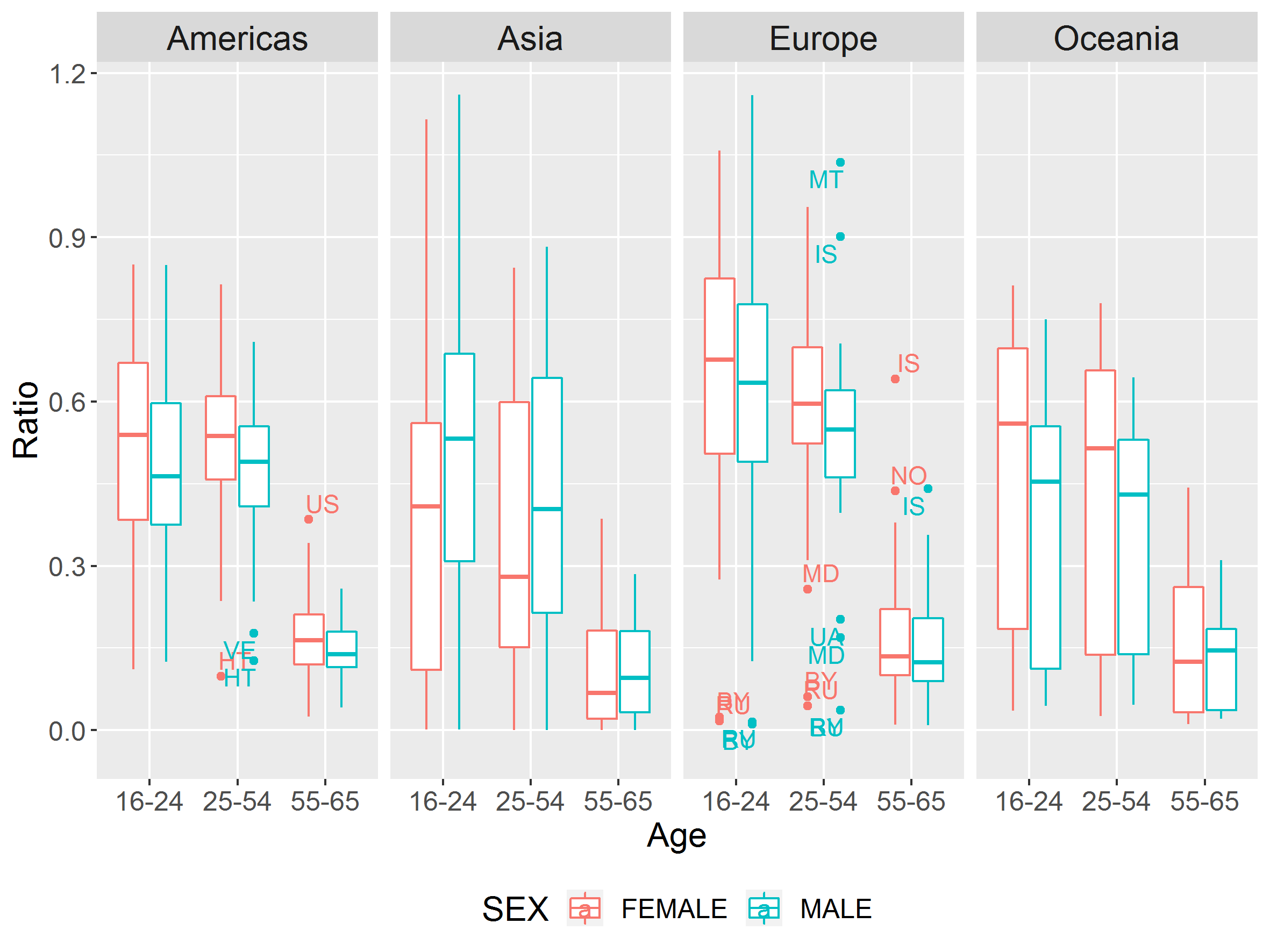}
    \caption{Box-plots of the Facebook penetration ratios of each country by gender and age}
    \label{Ratios}
    \end{center}
\end{figure}

We start by showing the FB penetration ratios (Eq.\ref{EqRatios}) for different age groups. From a cross-regional perspective, Europe is the continent with the lowest variability in FB ratios (Fig.\ref{Ratios}), whereas Asia is the continent with the highest one. The greatest median gender disparity in favor of the male population is in Asia. In North America, by contrast, the female population tend to be more engaged with FB than their male counterparts. This finding is consistent with previous work on gender gaps in FB usage~\cite{fatehkia_using_2018}. Here, we add the dimension of age, which was not considered in the previous literature. The addition of this variable led us to observe that in Asia, where  gender disparities in FB usage are particularly large, differences by gender in adoption rates disappear at older ages. This finding implies that gender inequality in online activity levels is mainly driven by differences at relatively young ages.  We cannot provide an explanation for this observation based on these data alone. In principle, it is possible that gender disparities decrease with age. However, we can also speculate that our observations could be the result of a  selection process, as older people who use FB may, on average, be better educated and have higher incomes than the general population, and thus have more equal access to the internet. Although this is only a hypothesis, we believe that it is important to point out that these effects could be driven by compositional changes in the population of online users. Estimates of gender disparities that do not account for these differences by age may lead to inaccurate representations of who has access to the internet and social media.  

\begin{figure}[ht]
    \begin{center}
    \includegraphics[width=.95\columnwidth]{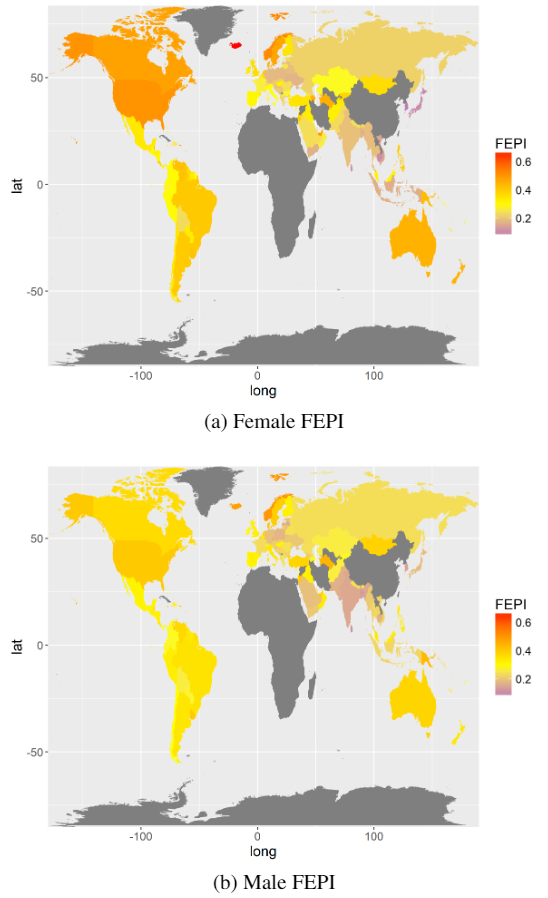}
    \caption{Maps of the Facebook Elderly Penetration Index (FEPI) by gender. The gray areas in the plot are countries that were not considered in this analysis.}
    \label{FEPI}
    \end{center}
\end{figure}

In order to understand the differences in the use of FB in middle and in late adulthood, we present values for the FEPI index (Eq.\ref{EqFEPI}) across countries. As Figure \ref{FEPI} shows, countries in  North America and northern Europe  have smaller gaps in the number of FB users in the middle and the late adulthood groups. This findings indicates that FB is adopted more equally across age groups.

\subsection{How the Size of the Social Network Varies by Age and by Gender}

In order to measure, albeit in an indirect way, the size of each user's friends network, we consider the variable \textit{close friends of people with birthdays in a month}, which gives the number of users who have friends with birthdays within a month from the date the data  were retrieved. More specifically, we evaluate the index presented in Equation \ref{EqIndex} for different age groups:
$$I(i,j)_{s,c}[\mbox{Close friends of people with birthdays in a month}]$$

\begin{figure}[ht]
    \begin{center}
    \includegraphics[width=.95\columnwidth]{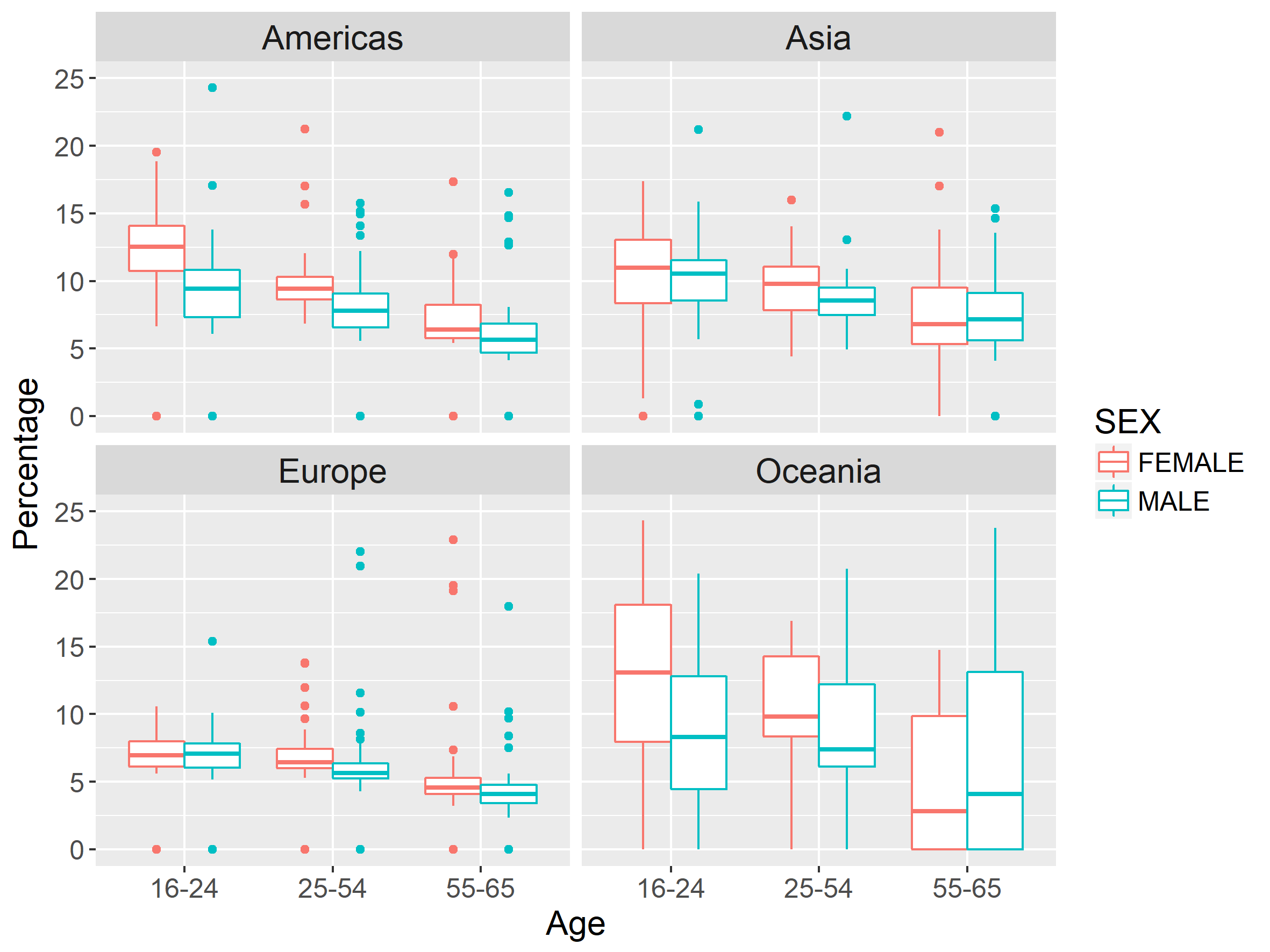}
    \caption{Box-plots of the median number of country users by age and gender who, according to FB, have close friends with birthdays in a month.}
    \label{Birthday}
    \end{center}
\end{figure}

As Figure \ref{Birthday} shows, women typically have a larger median number of close friends with birthdays in a month than men. The value is smaller for women than for men only among late-adulthood women in Asia and Oceania. In other words, women tend to have larger networks of close friends than men.

Across all regions, we observe that the fraction of users with close friends decreases with age. This may be related to the fact that there are fewer users at older ages. However, it is important to note that the decline by age is relatively small. This observation thus points to the possibility that FB plays an important role in helping older adults maintain friendships. In the context of population aging, the role of social media should be considered in discussions of issues like loneliness and social isolation. 

\subsection{How Facebook Adoption Varies across Demographic Groups `Away from Hometown'}

Social media may play a particularly important role in helping people who have moved, either internationally or domestically, maintain existing friendships or form new social bonds. For this reason, we looked at demographic differentials in FB adoption for the subgroup of users who are \textit{away from their hometown}. Concretely, we evaluated the index: 
$$I(i,j)_{s,c}[\mbox{Away from hometown}]$$

\begin{figure}[ht]
    \begin{center}
    \includegraphics[width=.95\columnwidth]{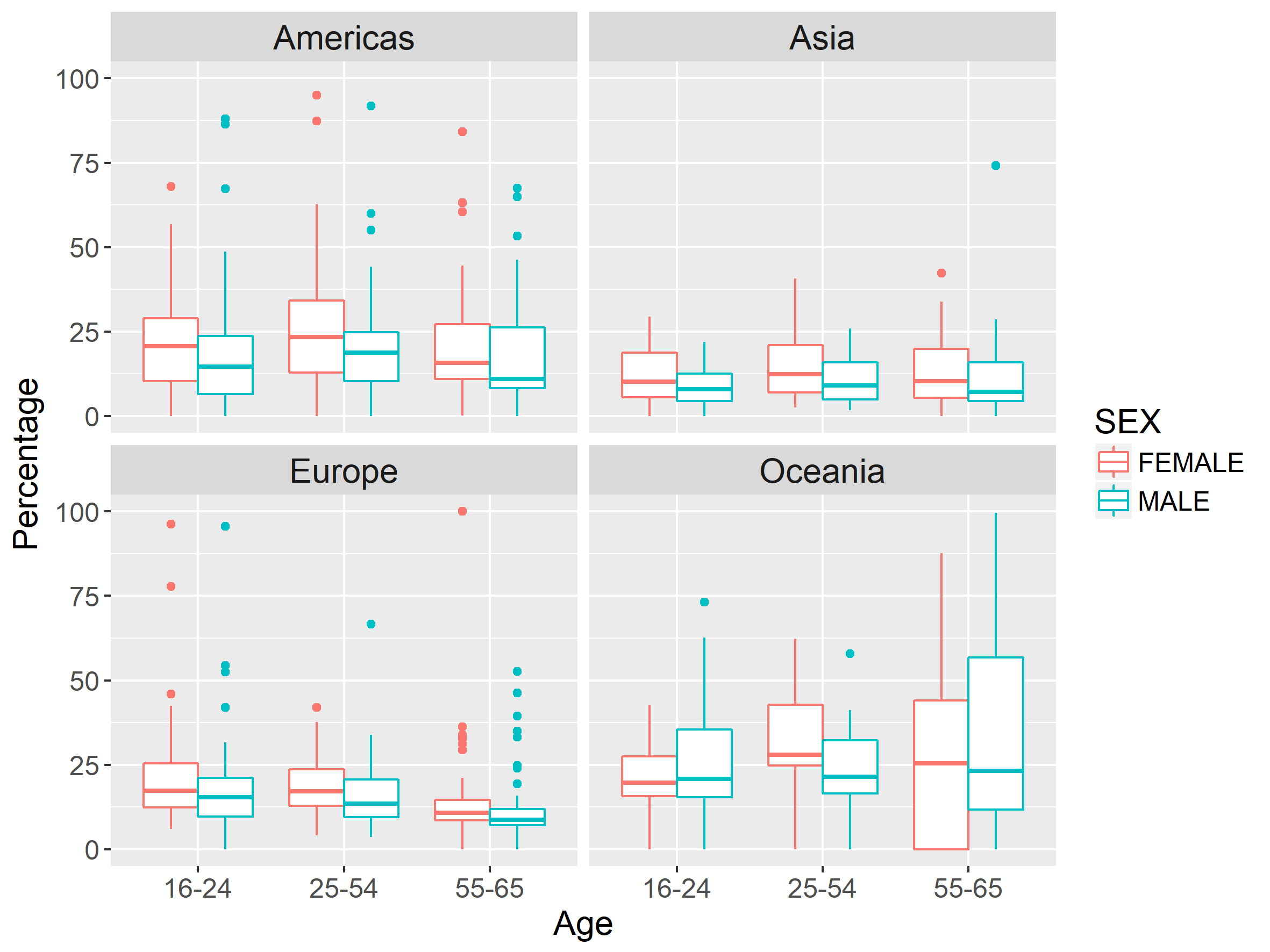}\\
    \caption{Box-plots of the median number of country users by age and gender who according to FB, are away from their hometown}
    \label{Away}
    \end{center}
\end{figure}

Figure \ref{Away} shows that the median number of female FB users classified as \textit{away from their hometown} by FB is greater than the median number of their male counterparts, regardless of their geographic region or age group. This finding indicates that when they are away from their hometown, women are more likely than men to engage with the FB platform, and that for this specific group of people which includes migrants we do not observe a gender gap in favor of men.

\section{Conclusions and Discussion}

Recent literature has highlighted the importance of evaluating digital gender gaps using online data, and particularly FB data for advertisers~\cite{fatehkia_using_2018,garcia_analyzing_2018}. In our paper, we contextualized those analyses and extended their scope by discussing the role of age, as well as broader patterns of demographic differentials in FB adoption around the world.  

We found that in countries in North America and northern Europe, patterns of FB adoption do not differ greatly between older and younger adults. In Asian countries, which have high levels of gender inequality, differences in levels of FB adoption by gender disappear at older ages, possibly because of selection processes in the socioeconomic characteristics of users from different demographic groups. We also observed that across countries, women tend to have larger networks of close friends than men, and that female users who are away from their hometown are more likely to engage in FB use than their male counterparts, regardless of their region and age group. 

Our findings are descriptive in nature, but have important implications. We showed that gender gaps in the usage of digital platforms, and of FB in particular, cannot be fully understood without accounting for key demographic variables like age. Importantly, we also highlighted that in addition to differences in the prevalence of FB use, there are important qualitative differences in how the platform is used. While the size of the network of close friends typically decreases with age, women tend to maintain larger networks than men, and tend to be engaged with FB at higher rates if they are away from their hometown (and potentially away from their network of extended family). As our findings are intended to be a first step towards gaining a more comprehensive understanding of demographic differentials in access to digital technologies and their implications for our societies, further work needs to be done in order to validate the results. Validation could be achieved by using representative surveys like SHARE\footnote{\url{http://www.share-project.org/home0.html}} or running surveys in FB.

\bibliographystyle{aaai}
\bibliography{PP-GilClavelS.1072}{} 
\end{document}